# Reduction of Nitrogen Oxides (NO$_x$) by Superalkalis

Ambrish Kumar Srivastava

P. G. Department of Physics, Veer Kunwar Singh University, Ara-802301, Bihar, India

E-mail: ambrishphysics@gmail.com

**Abstract**

$NO_x$ are major air pollutants, having negative impact on environment and consequently, human health. We propose here the single-electron reduction of $NO_x$ ($x$ = 1, 2) using superalkalis. We study the interaction of $NO_x$ with $FLi_2$, $OLi_3$ and $NLi_4$ superalkalis using density functional calculations, which lead to stable superalkali-$NO_x$ ionic complexes with negatively charged $NO_x$. This clearly reveals that the $NO_x$ can successfully be reduced to $NO_x^-$ anion due to electron transfer from superalkalis. It has been also noticed that the size of superalkalis plays a crucial in the single-electron reduction of $NO_x$.

## 1. Introduction

Nitrogen oxides (NO and $NO_2$) are known to be major air pollutants, collectively referred to as $NO_x$. They contribute to the pollution by forming smog, acid rain as well as tropospheric ozone. These are generally produced by reaction of nitrogen and oxygen during combustion of vehicle fuels [1, 2] as well as by lightening during thunderstorms [3, 4]. Apart from pollution, $NO_x$ gases have significant impact on human health indirectly. For instance, $NO_x$ gases react with certain organic compounds forming smog and destroying ozone, which possesses adverse health effects such as damage to lung tissue and reduction in lung function [5]. For the sake of environmental safety, therefore, it is desirable to lessen the content of $NO_x$ in atmosphere. This can be achieved by using the reductants such as urea or ammonia with or without use of a catalyst. In such reduction processes, $NO_x$ are generally converted into nitrogen molecule ($N_2$), water ($H_2O$) and carbon dioxide ($CO_2$).

The single-electron reduction of $NO_x$ has been difficult, particularly of NO due to its negligible electron affinity, 0.026 eV [6]. Although few $NO_2$ salts such as $NaNO_2$ already exist [7], there is no corresponding salt of NO. However, recent studies [8, 9] suggest that $CO_2$ can be easily reduced to $CO_2^-$ anion by using superalkalis, despite the fact that $CO_2$ possesses no positive electron affinity [10]. This prompted us to enquire whether superalkalis can be employed for single-electron reduction of $NO_x$. Superalkalis [11] are species whose ionization energies (IEs) are lower than those of halogen. Such species have been previously employed in the design of supersalts [12-15], superbases [16-19], alkalides [20-22], etc. In general, superalkalis are hypervalent species with the general formula $XM_{k+1}$ (where $k$ is the valence of electropositive atom M and X is an electronegative ligand) possessing excess electron, which offer them strong reducing capabilities. It is, therefore, instructive to observe whether superalkalis can be employed to transfer an electron to $NO_x$. In this letter, we report the interaction of $NO_x$ with superalkalis using density functional theory based method. We noticed that superalkalis can indeed be used to reduce $NO_x$ into $NO_x^-$ anion. The study paves a way to reduce major air pollutants, which might be useful from the perspectives of environmental safety and human health as well.

## 2. Computational details

All computations were performed using density functional theory with B3LYP [23, 24] hybrid functional and 6-311+G(d) basis set in Gaussian 09 program [25]. Considering the size of system, it is quite easy to employ high level *ab initio* methods such as the second order Møller-Plesset perturbative (MP2) theory [26], the use of B3LYP is due to the fact that it is capable to reproduce the experimental results. In order to confirm this, we have performed test calculations on $NO_x$ and compared the results in Table 1. One can note that our B3LYP computed bond lengths of $NO_x$ reproduce corresponding experimental values [27], unlike MP2 values. Furthermore, B3LYP computed electron affinity of $NO_2$ is very close to corresponding experimental value measured by photoelectron spectrum [28] (see Table 1). This explains our preference of B3LYP over MP2 in the present study.

The equilibrium geometries have been obtained without any symmetry constraints in the potential energy surface. The vibrational analysis has been performed at the same level of theory to ensure that all frequency values are positive, i.e., the structures belong to true minima in the potential energy surface. The binding energy (BE) of $SA-NO_x$ complexes are calculated by using following equations;

$BE = E[NO_x] + E[SA] – E[SA-NO_x]$

where $E[..]$ represents total (electronic + zero-point) energy of respective species. The charge transfer in SA-NO$_x$ complexes has been computed using natural population analysis (NPA) scheme [29] using NBO program [30] as implemented in Gaussian 09.

## 3. Results

First we analyze the equilibrium structures of NO$_x$ and SA systems as displayed in Fig. 1. We have chosen FLi$_2$, OLi$_3$ and NLi$_4$ superalkalis, which belong to the general formula of XM$_{k+1}$ such that $k = 1$ for F, 2 for O and 3 for N. The equilibrium structures of FLi$_2$, OLi$_3$ and NLi$_4$ are bent, trigonal planar and tetrahedral, respectively. In cationic form, the structure of FLi$_2$ becomes linear and the bond lengths of OLi$_3$ and NLi$_4$ are slightly increased. The IEs of these species are calculated to be 4.20 eV for FLi$_2$, 3.85 eV for OLi$_3$ and 3.75 eV for NLi$_4$, smaller than that of Li (~5 eV) as expected. In case of NO$_x$, the addition of an extra electron, i.e., single-electron reduction increases the bond length by ~0.1 Å. The bond angle of NO$_2^-$ is also reduced to 117$^o$ from 134$^o$ in NO$_2$. The negative charge on NO$_2^-$ is equally shared by two O atoms. Below we discuss the single-electron reduction of NO$_x$ by interaction with SA species. We have considered various possible positions of NO$_x$ with respect to SA species.

The equilibrium structures of SA-NO complexes are displayed in Fig. 2 and corresponding parameters are listed in Table 2. For each SA species, we obtain two possible low lying isomers of SA-NO complexes (with the energy difference of merely 0.02 eV) such that either N or both O and N interact with SA. The lowest energy corresponds to the structure in which both O and N atoms interact with two Li atoms of SAs. In all SA-NO complexes, the bond lengths N−Li and O−Li in the lowest energy isomers (a) range 2.037−2.074 Å and 1.857−1.882 Å, respectively. Furthermore, the bond length of NO is increased to ~1.250 Å, which become comparable to that of NO$^-$ (see Fig. 1). This may indicate that the charge transfer takes place from SA to NO in SA-NO complexes. The BE of SA-NO complexes are listed in Table 2. For all SA-NO complexes, BE > 0 and therefore, all these complexes are energetically stable. One can also note that the BE of SA-NO complexes decreases, BE(FLi$_2$-NO) > BE(OLi$_3$-NO) > BE(NLi$_4$-NO), with the increase in the size of SAs (FLi$_2$ < OLi$_3$ < NLi$_4$). Thus, the interaction between superalkali and NO becomes slightly weaker with the increase in the size of superalkalis. In order to analyze the charge transfer in SA-NO complexes, we have computed NPA charges ($\Delta q$) on NO moiety as listed in Table 2. In SA-NO complexes, the $\Delta q$ takes value of -0.78$e$ for SA = FLi$_2$, -0.76$e$ for OLi$_3$ and -0.72$e$ for NLi$_4$. These $\Delta q$ values follow the same trend as those of BE values as expected due to the fact that lower $\Delta q$ value results in the decrease in the strength of charge-transfer interaction. Thus, the charge transfer to NO is close to unity, which clearly suggests that the SA species are capable to reduce NO to NO⁻ and form stable ionic compounds, namely, (SA)$^+$(NO)$^-$.

In Fig. 3, we have displayed the equilibrium structures of SA-NO$_2$ complexes and corresponding parameters are collected in Table 3. We obtain two possible isomers, (a) and (b) of FLi$_2$-NO$_2$ in which NO$_2$ binds with two Li-atoms and one Li-atom, respectively. In case of OLi$_3$-NO$_2$ and NLi$_4$-NO$_2$, however, we obtain three isomers, (a), (b) and (c) whose relative energies are listed in Table 3. In each case, the lowest energy of SA-NO$_2$ complexes corresponds to the structure in which both O atoms of NO$_2$ interact with two Li-atoms of superalkalis as displayed in Fig. 3. The BE of SA-NO$_2$ complexes are also listed in Table 3. All BE values are positive except that of isomer (c) of NLi$_4$-NO$_2$. Furthermore, these BE values (~3-5 eV) are larger than those of corresponding SA-NO complexes (~1-2 eV). This may suggest that the interaction of NO$_2$ with SA is stronger than that of NO. This can be expected due to larger EA of NO$_2$ as compared to NO (see Table 1). Like SA-NO complexes, however, the BEs of SA-NO$_2$ complexes follow the same trend as: BE(FLi$_2$-NO$_2$) > BE(OLi$_3$-NO$_2$) > BE(NLi$_4$-NO$_2$). Nevertheless, the charge transfer to NO$_2$ ($\Delta q$) is increased as compared to NO (see Table 2) in accordance with their BE values, becoming very close to unity. Likewise, SA-NO$_2$ complexes can also be expressed as (SA)$^+$(NO$_2$)$^-$.

## 4. Discussion

The results described above clearly suggest that the SA species are indeed capable to reduce NO$_x$ to corresponding anions and form stable SA-NO$_x$ ionic complex. However, the electron from SA can be easily transferred to NO$_2$ than to NO due to larger EA of the former. Consequently, SA-NO$_2$ complexes are more stable as compared to SA-NO complexes, which is reflected in their BE values. The size of superalkalis (SAs) and their IEs are two important parameters for reduction of NO$_x$. It should, however, be noticed that the size of superalkalis are more effective than their ionization energy in NO$_x$-reduction. For example, the size of superalkalis considered are in the order, FLi$_2$ < OLi$_3$ < NLi$_4$ whereas their ionization energies follow a reverse order, FLi$_2$ > OLi$_3$ > NLi$_4$. The charge transfer to NO$_x$ is in the order, $\Delta q$(FLi$_2$-NO$_x$) > $\Delta q$(OLi$_3$-NO$_x$) > $\Delta q$(NLi$_4$-NO$_x$). This is consistent with the fact that larger superalkalis usually possess more delocalized electron cloud [13]. Therefore, charge transfer from larger superalkali such as NLi$_4$ is not as much effective as in case of smaller superalkali such as FLi$_2$. Therefore, FLi$_2$ should be considered as the most powerful in the single-electron reduction of NO$_x$. The similar conclusion has already been drawn in case of CO$_2$-reduction by some novel superalkalis [8].

In order to further explore this fact, we have also carried out similar calculations using ONa$_3$ superalkali. The IE of ONa$_3$ is computed to be 3.71 eV, which is smaller than that of OLi$_3$ (3.85 eV). The equilibrium structures of ONa$_3$-NO$_x$ complexes are displayed in Fig. 4. One can see that unlike $OLi_3$-NO, the lowest energy of $ONa_3$-NO corresponds to the structure in which N binds with two Na atoms. Similarly, the isomers of $ONa_3$-$NO_2$ are slightly different than the isomers of $OLi_3$-$NO_2$. However, the charge transfer to $NO_x$ from $ONa_3$ superalkali becomes -0.75$e$ for x = 1 and -0.80$e$ for x = 2. This may further support the fact that the size of superalkalis is more crucial in the reduction of $NO_x$. Note that the crystal structure of $Na_3NO_3$ has already been reported by Jansen [31], which was obtained by heating $Na_2O$ and $NaNO_2$. According to Jansen, $Na_3NO_3$ is not an orthonitrite but it contains $NO_2^-$ anionic group. This may not only support our conclusion that $ONa_3$ superalkali is capable in reducing $NO_2$ to $NO_2^-$ but also suggest that these species are feasible experimentally, at least in the gas phase.

## 5. Conclusions

Using density functional B3LYP/6-311+G(d) calculations, we have studied the interaction of NO and $NO_2$ with typical superalkalis such as $FLi_2$, $OLi_3$ and $NLi_4$. We have obtained the lowest energy structures of resulting superalkali-$NO_x$ ($x$ = 1 and 2) complexes along with possible isomers. These complexes are ionic and stable in which an electron transfer takes place from superalkali to $NO_x$ moieties. Thus, $NO_x$ can be successfully reduced to its anion by using superalkalis. We have also noticed that the single-electron reduction of $NO_x$ is more effective in case of smaller superalkali, similar to the case of $CO_2$-reduction reported previously. We have also studied the reduction of $NO_x$ using $ONa_3$ superalkali forming $ONa_3$-NO and $ONa_3$-$NO_2$ complexes. This finding is supported by the fact that $ONa_3$-$NO_2$ ionic complex has already been synthesized.

## Acknowledgement

A. K. Srivastava acknowledges Prof. N. Misra, Department of Physics, University of Lucknow for providing computational facilities and helpful suggestions.

**References**

[1] M. R. Beychok, Oil Gas J. 1973 (1973 53-56.

[2] H. Omidvarborna, A. Kumar, D. -S. Kim, Fuel Process. Technol. 140 (2015) 113-118

[3] J. S. Levine, T. R. Augustsson, I. C. Andersont, J. M. Hoell Jr., D. A. Brewer, Atmos. Environ. 18 (1984) 1797–1804.

[4] U. Schumann, H. Huntrieser, Atmos. Chem. Phys. 7 (2007) 3823.

[5] NOx How Nitrogen Oxides Affect The Way We Live And Breathe, Available online at: https://nepis.epa.gov/Exe/ZyPURL.cgi?Dockey=P10006ZO.txt (Retrieved 20-12-2017).

[6] M. J. Travers, D. C. Cowles, G. B. Ellison, Chem. Phys. Lett. 164 (1989) 449.

[7] N. N. Greenwood, A. Earnshaw, Chemistry of the Elements (2$^{nd}$ ed.), Butterworth-Heinemann (1997).

[8] T. Zhao, Q. Wang, P. Jena, Nanoscale 9 (2017) 4891-4897.

[9] H. Park, G. Meloni, Dalton Trans. 6 (2017) 11942-11949.

[10] A. Knapp, O. Echt, D. Kreisle, T. D. Mark, E.Recknagel, Chem. Phys. Lett.126 (1986) 225.

[11] G. L. Gutsev, A. I. Boldyrev, Chem. Phys. Lett. 92 (1982) 262.

[12] Y. Li, D. Wu, Z.-R. Li, Inorg. Chem. 47 (2008) 9773-9778.

[13] H. Yang, Y. Li, D. Wu, Z.-R. Li, Int. J. Quantum Chem. 112 (2012) 770-778.

[14] A. K. Srivastava, N. Misra, Mol. Phys. 112 (2014) 2621-2626.

[15] Y. -Q. Jing, Z. -R. Li, D. Wu, Y. Li, B. -Q. Wang, F. L. Gu, Y. Aoki, ChemPhysChem 7 (2006) 1759–1763.

[16] A. K. Srivastava, N. Misra, New J. Chem. 39 (2015) 6787-6790.

[17] A. K. Srivastava, N. Misra, RSC Adv. 5 (2015) 74206-74211.

[18] A. K. Srivastava, N. Misra, Chem. Phys. Lett. 648 (2016) 152-155.
9

Table 1. Comparison of calculated and experimental values for NO$_x$.

| Parameter | NO | | | NO$_2$ | | |
|---|---|---|---|---|---|---|
| | MP2 | B3LYP | Expt. | MP2 | B3LYP | Expt. |
| Bond length (Å) | 1.135 | 1.148 | 1.151[a] | 1.203 | 1.193 | 1.193[a] |
| Bond angle (°) | - | - | - | 133.9 | 134.4 | 134.1[a] |
| Electron affinity (eV) | -0.33 | 0.36 | 0.026[b] | 1.88 | 2.29 | 2.273[c] |

a) Ref. [27]
b) Ref. [6]
c) Ref. [28]

Table 2. B3LYP/6-311+G(d) calculated structures, their relative energy (ΔE), binding energy (BE) and charge located on NO (Δq) of SA-NO complexes.

| Complex | Isomer | Symmetry | ΔE (eV) | BE (eV) | Δq (e) |
|---|---|---|---|---|---|
| $FLi_2$-NO | a | $C_s$ | 0 | 1.52 | -0.78 |
| | b | $C_{2v}$ | 0.02 | 1.50 | -0.78 |
| $OLi_3$-NO | a | $C_s$ | 0 | 1.30 | -0.76 |
| | b | $C_{2v}$ | 0.02 | 1.28 | -0.76 |
| $NLi_4$-NO | a | $C_s$ | 0 | 1.04 | -0.72 |
| | b | $C_{2v}$ | 0.02 | 1.02 | -0.72 |

Table 3. B3LYP/6-311+G(d) calculated structures, their relative energy (Δ$E$), binding energy (BE) and charge located on NO (Δ$q$) of SA-NO$_2$ complexes.

| Complex | Isomer | Symmetry | Δ$E$ (eV) | BE (eV) | Δ$q$ ($e$) |
| --- | --- | --- | --- | --- | --- |
| FLi$_2$-NO$_2$ | a | C$_{2v}$ | 0 | 4.45 | -0.84 |
| | b | C$_{2v}$ | 0.83 | 3.62 | -0.83 |
| OLi$_3$-NO$_2$ | a | C$_{2v}$ | 0 | 4.26 | -0.81 |
| | b | C$_s$ | 0.30 | 3.96 | -0.81 |
| | c | C$_1$ | 0.89 | 3.37 | -0.80 |
| NLi$_4$-NO$_2$ | a | C$_{2v}$ | 0 | 4.00 | -0.78 |
| | b | C$_1$ | 0.23 | 3.77 | -0.78 |
| | c | C$_2$ | 5.21 | -1.21 | -0.79 |

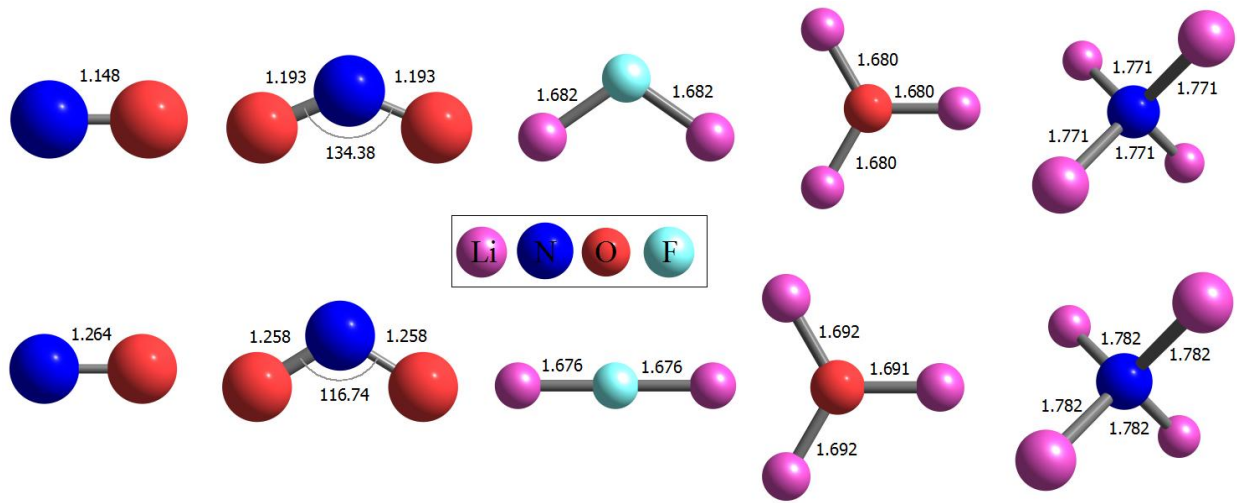

Fig. 1. Equilibrium structure of NO, $NO_2$ (upper row) and their anions (lower row). The equilibrium structure of superalkalis (upper row) and their cations (lower row) are also displayed. All bond lengths are given in Å.

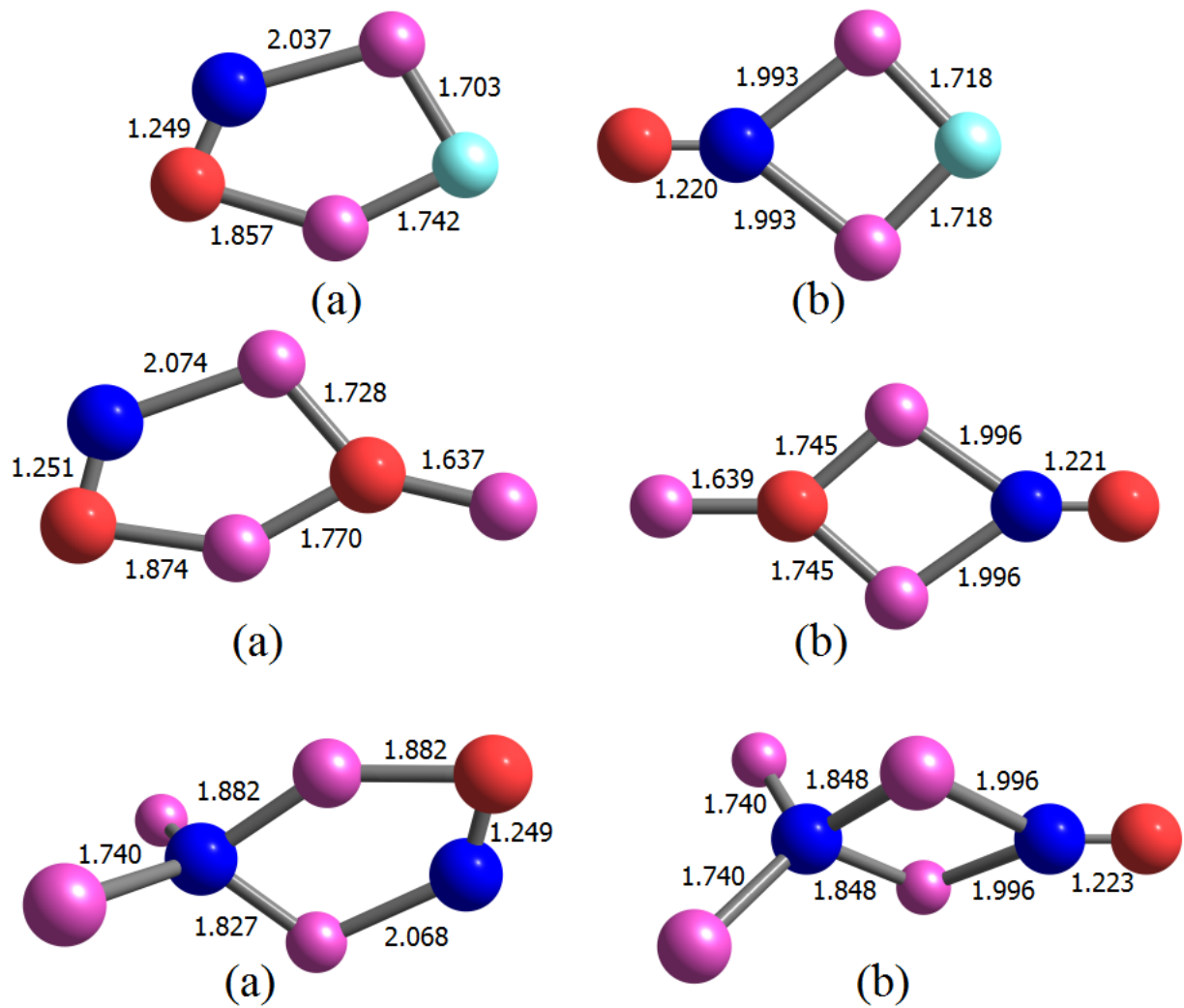

Fig. 2. Equilibrium structures of SA-NO complexes and their isomers with bond lengths in Å.

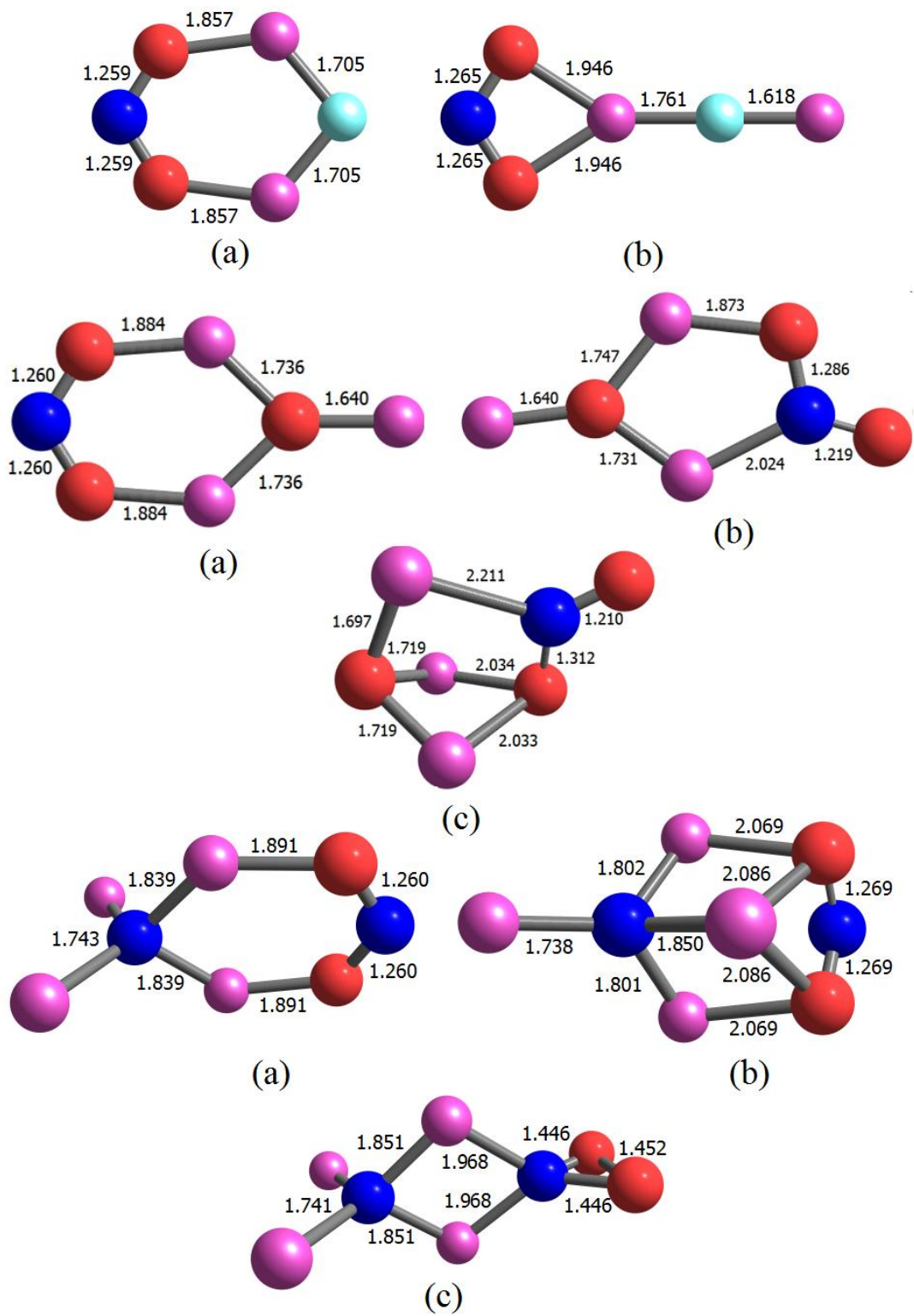

Fig. 3. Equilibrium structures of SA-NO$_2$ complexes and their isomers with bond lengths in Å.

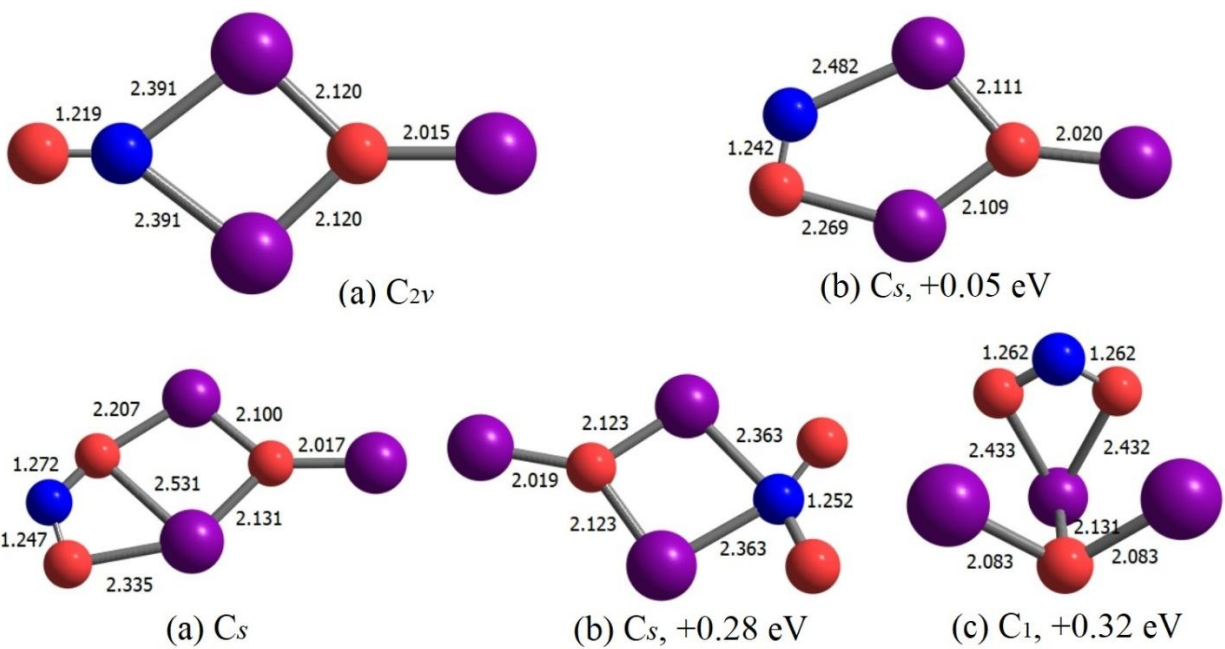

Fig. 4. Equilibrium structures of ONa$_3$-NO and ONa$_3$-NO$_2$ complexes and their isomers with bond lengths in Å. Symmetry and relative energy of isomers are also displayed.